\documentclass[%
reprint,
superscriptaddress,
frontmatterverbose,
amsmath,amssymb,
pra,
]{revtex4-2}
\usepackage{amsfonts,amsmath,amssymb,amsthm}
\usepackage[normalem]{ulem}

\usepackage{graphicx}% Include figure files
\usepackage{dcolumn}% Align table columns on decimal point
\usepackage{bm}% bold math

\usepackage{csquotes}
\MakeOuterQuote{"}

\usepackage{hyperref}

\hypersetup{
	pdfborder={0 0 0}
	pdfnewwindow=true,      % links in new window
	colorlinks=true,       % false: boxed links; true: colored links
	linkcolor=blue,          % color of internal links (change box color with linkbordercolor)
	citecolor=blue,        % color of links to bibliography
	filecolor=blue,      % color of file links
	urlcolor=blue,          % color of external links
}

\usepackage{color}
\usepackage{amsmath}

\usepackage{lineno}
\makeatother

\newcommand{\be}{\begin{equation}}
\newcommand{\ee}{\end{equation}}

\def\ket#1{\mathinner{|{#1}\rangle}}

\usepackage{braket}
\usepackage{array}
\usepackage{booktabs}
\usepackage{multirow}

\usepackage{float}

\newif\ifdraft
\drafttrue
\draftfalse
\usepackage{titlesec}
\titleformat{\section}{\raggedright\bfseries}{\arabic{section}.}{1em}{}

\begin{document}

% \preprint{APS/123-QED}

\title{Neural-Quantum-States Impurity Solver for Quantum Embedding Problems}
\author{Zhanghao Zhouyin}
% \email{yinzhanghao.zhou@mail.mcgill.ca}
% \thanks{These authors contributed equally.}

\affiliation{Department of Physics, McGill University, Montreal, Quebec, Canada H3A2T8}

\author{Tsung-Han Lee}
\affiliation{Department of Physics, National Chung Cheng University, Chiayi 62102, Taiwan}

\author{Ao Chen}
\affiliation{Center for Computational Quantum Physics, Flatiron Institute, New York, New York 10010, USA}
\affiliation{Division of Chemistry and Chemical Engineering,
California Institute of Technology, Pasadena, California 91125, USA}
\author{Nicola Lanatà}
\affiliation{School of Physics and Astronomy, Rochester Institute of Technology, Rochester, New York 14623, USA}
\affiliation{Center for Computational Quantum Physics, Flatiron Institute, New York, New York 10010, USA}

\author{Hong Guo}
% \altaffiliation{test}
%\email{hong.guo@mcgill.ca}
\affiliation{Department of Physics, McGill University, Montreal, Quebec, Canada H3A2T8}

% \author{Weinan  E}
% %\email{weinan@math.pku.edu.cn}
% \affiliation{AI for Science Institute, Beijing 100080, China}
% \affiliation{School of Mathematical Science, Peking University, Beijing 100871, China}
% \affiliation{Center for Machine Learning Research, Peking University, Beijing, 100871 China}

\date{\today}

\begin{abstract}

Neural quantum states (NQS) have emerged as a promising approach to solve second-quantized Hamiltonians, because of their scalability and flexibility. In this work, we design and benchmark an NQS impurity solver for the quantum embedding (QE) methods, focusing on the ghost Gutzwiller Approximation (gGA) framework. We introduce a graph transformer-based NQS framework able to represent arbitrarily connected impurity orbitals of the embedding Hamiltonian (EH) and develop an error control mechanism to stabilize iterative updates throughout the QE loops. We validate the accuracy of our approach with benchmark gGA calculations of the Anderson Lattice Model, yielding results in excellent agreement with the exact diagonalisation impurity solver. 
Finally, our analysis of the computational budget reveals the method's principal bottleneck to be the high-accuracy sampling of physical observables required by the embedding loop, rather than the NQS variational optimization, directly highlighting the critical need for more efficient inference techniques.
\end{abstract}

\maketitle

\section{Introduction}

Materials with strong electronic correlations have been the subject of extensive investigation in physics and materials science for decades. Such systems exhibit a variety of electronic phases, including metallic, insulating, and superconducting states~\cite{mahan2013many}, and represent a versatile design space for technological applications in electronics, quantum computing, and sensing. Designing new correlated materials with targeted properties therefore depends on the ability to solve the many-body electronic Hamiltonian, a computationally demanding task~\cite{adler2018correlated}. 
QE methods provide a robust framework for overcoming these challenges \cite{sun2016quantum, ma2021quantum, knizia2013density}. The common strategy underlying these methods is to describe each fragment of interacting orbitals through an effective EH, where its complex environment is replaced by a simpler, entangled quantum bath, designed to approximate its influence~\cite{sun2016quantum, mejuto2023efficient}. 
The link between this effective model and the original system is established through self-consistency conditions, and different embedding schemes are defined by their choice of which physical property to match~\cite{sun2016quantum}.
Dynamic mean-field theory (DMFT) 
~\cite{dmft_book,Anisimov_DMFT,Held-review-DMFT,georges1996dynamical,xidai_impl_LDA+DMFT,kotliar2006electronic,savrasov2001correlated, paul2019applications}, for instance, uses frequency-dependent one-body Green's functions, whereas density matrix embedding theory (DMET) \cite{knizia2013density, knizia2012density, hermes2020variational, wouters2016practical, sekaran2021householder, cui2019efficient} typically uses one- and two-body density matrices. 
The recently developed gGA~\cite{lanata2017emergent, frank2021quantum, lee2024charge, lanata2023derivation, lanata2022operatorial, lee2023accuracy, lee2023accm} is a powerful variational method that generalizes the standard Gutzwiller Approximation (GA)~\cite{Gutzwiller3,Our-PRX,lanata-barone-fabrizio,Fang,Ho,Gmethod,Bunemann,Attaccalite} by systematically extending its variational space with auxiliary "ghost" fermionic degrees of freedom. This approach yields results in remarkable agreement with DMFT but at a much lower computational cost, as it requires calculating only the ground state of a finite-size impurity model, whereas DMFT requires the full spectra from an impurity model that corresponds to an infinite bath.
Successful applications of gGA include accurate modeling of the Anderson lattice systems \cite{frank2021quantum}, excitonic phenomena \cite{lanata2017emergent}, non-equilibrium systems \cite{guerci2023time} and altermagnetic systems \cite{giuli2025altermagnetism}, along with extensions that achieve charge self-consistency with density functional theory (DFT) \cite{lee2024charge}, demonstrating its versatility and practical utility in real-material contexts.

\begin{figure}[htbp!]
\includegraphics[width=8.5 cm]{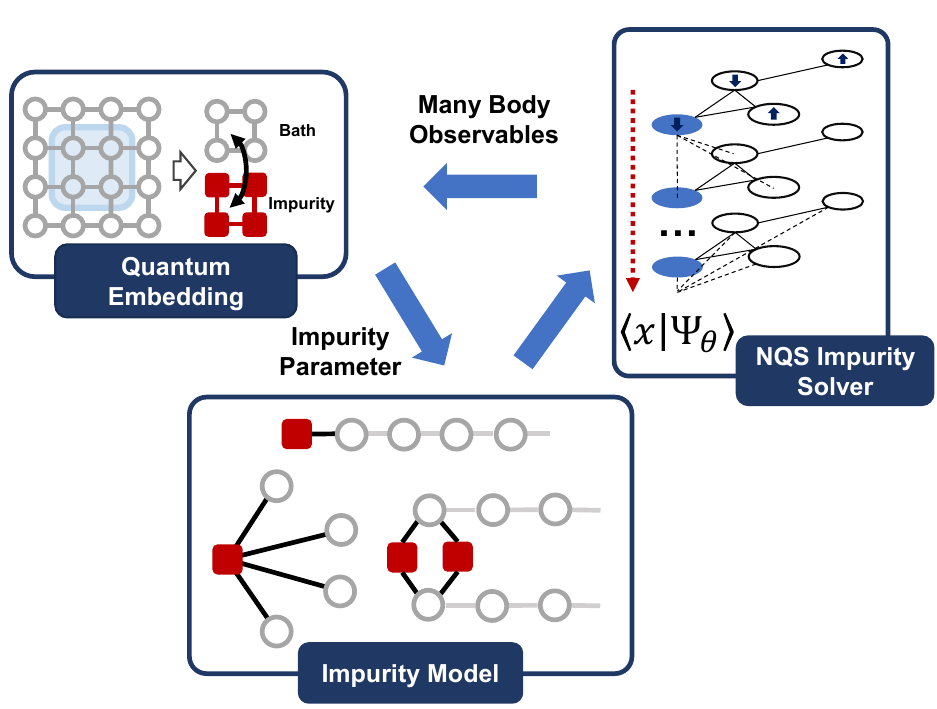}
\caption{The neural quantum states - quantum embedding (NQS-QE) workflow. The many-body Hamiltonian is embedded into impurity models via QE algorithms. It is then solved by the NQS impurity solver. The observables are derived from the graph neural network representation of the many-body wave function, which is then used to update the impurity parameters, thereby forming a self-consistent loop.}
\label{fig:model}
\end{figure} 

Despite the advances offered by QE methods, solving the embedding impurity Hamiltonian model remains a central computational bottleneck. Current impurity solvers used with QE include exact diagonalization (ED) \cite{caffarel1994exact, capone2007solving}, Numerical Renormalization Group (NRG) \cite{wilson1975renormalization, bulla2008numerical}, Density Matrix Renormalization Group (DMRG) \cite{white1992density, schollwock2005density}, Natural Orbital Renormalization Group\cite{wang2022solving,wang2025ab}, Tensor Network \cite{chen2024grassmann,chen2024regrassmann} and Quantum Monte Carlo (QMC) \cite{gull2011continuous, rubtsov2005continuous}. While each of these methods has distinct advantages, their computational complexity often limits their application to practically relevant systems.

Neural Quantum States (NQS) \cite{carleo2017solving, sharir2022neural, chen2024empowering, doi:10.1073/pnas.2122059119, GOLDSHLAGER2024113351, levine2019quantum,yu2024solving}, due to their inherent flexibility in representing wave functions in second-quantized formalisms, offer a compelling alternative to some of these challenges. In particular, their variational foundation provides us with controlled approximation errors through parameter optimization \cite{lange2024architectures}. Additionally, variational Monte Carlo provides a favorable computational scaling \cite{yokoyama1987variational} especially for multi-orbital scenarios. Specifically in this work, we adopted the recently proposed hidden pfaffian ansatz, which has a proven scaling of $O(N^2)\sim O(N^3)$ \cite{chen2025neural} with N the number of sites. The rapidly developing field of NQS shows promising trends toward a unified and scalable impurity solver \cite{cao2024vision}. Recent works \cite{ma2024quantum} incorporated the QiankunNet NQS method with DMET and applied successfully to calculate the ground state properties of atomic systems. Nonetheless, impurity solvers based on NQS must address specific requirements such as handling irregular hopping patterns that demand arbitrary orbital interactions and ensuring numerical stability and consistency, which are required for a robust solver embedded in the QE self-consistency workflow. Accurately calculating observables from the optimized wave functions further introduces additional methodological challenges.

In this work, we develop a graph transformer-based NQS impurity solver for the quantum EH, which is able to represent arbitrarily connected impurity orbitals. We implement a systematic error control framework to rigorously bound uncertainties arising during the iterative optimization and embedded properties sampling of the NQS wave function. Our model initializes node features with spin configurations and employs iterative graph attention mechanisms, achieving high flexibility in wave function representation. Optimization efficiency is enhanced using recently proposed algorithms such as the Neural Network Pfaffian \cite{chen2025neural}, MinSR\cite{chen2024empowering} and SPRING\cite{GOLDSHLAGER2024113351}. Coupling our graph-based NQS solver with the gGA method, we validate its performance using benchmark Anderson Lattice Models \cite{frank2021quantum}, obtaining excellent agreement with ED solutions. Our results demonstrate the feasibility of employing NQS as an impurity solvers within QE algorithms. Furthermore, we conduct extensive analyses of convergence behaviors across different error control parameters and computational timing, highlighting critical open questions that must be addressed to achieve scalable and practically relevant NQS-based impurity solvers.

The rest of the paper is organized as follows. In the next section, the technical details of the method are presented, including the NQS impurity solver, its associated error control, and its integration with the gGA QE approach. In section 3, benchmark results associated with the Anderson Lattice Model and a detailed cost and error analysis of the NQS solver are illustrated. Finally, section 4 presents a discussion and a summary of the NQS-gGA many-body formalism developed in this work. 

\begin{figure}[htbp!]
\includegraphics[width=9.5 cm]{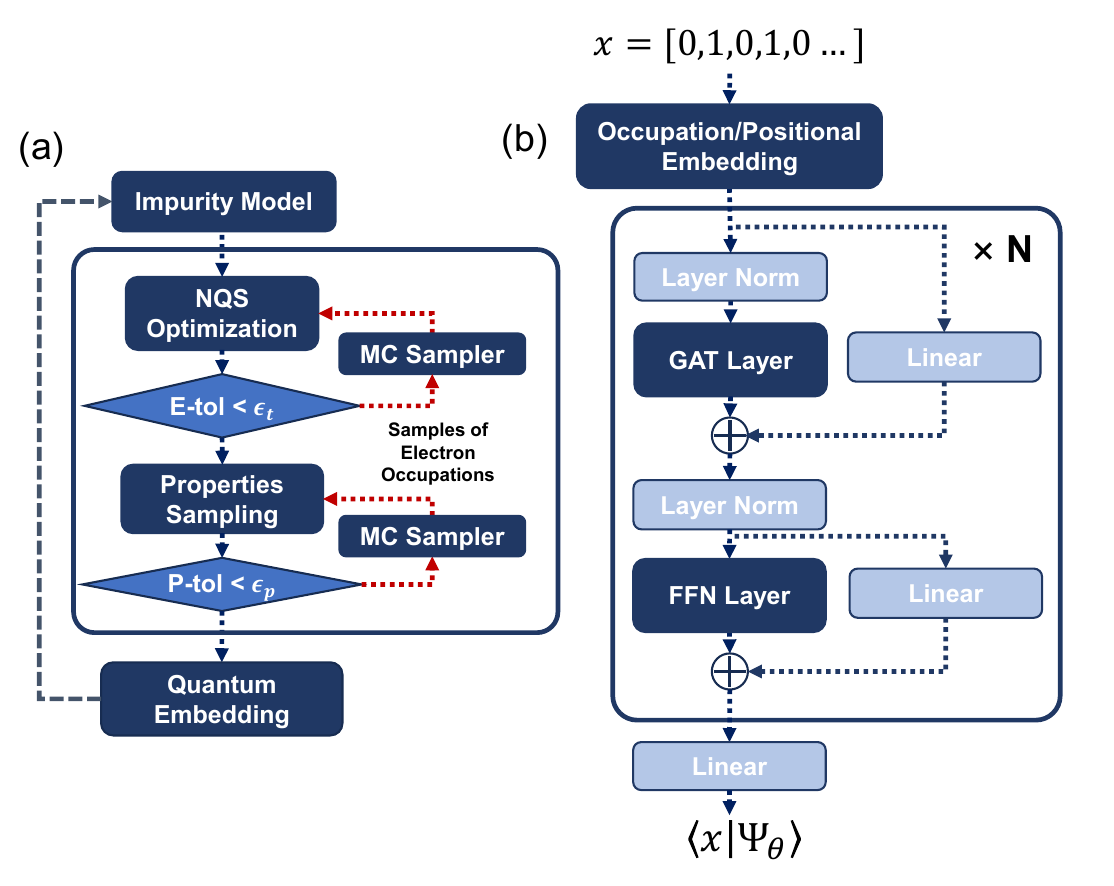}
\caption{(a) The workflow for the neural quantum states (NQS) impurity solver. The outer loop shows the impurity solver's coupling with the QE method, which generates an impurity model for the NQS to solve. The solving process includes the NQS optimization and properties sampling steps, with error control metrics E-tol and P-tol to guarantee convergence. The E-tol and P-tol criteria are checked against user-defined tolerances $\epsilon_t$ and $\epsilon_p$ for the optimization and sampling steps, respectively. Both steps are driven by the Monte Carlo (MC) sampler to compute expectations. (b) The neural network wave function's architecture. The spin orbital configuration is first encoded with the occupation and positional embedding method, and passes through the main body of the network, which comprises \textbf{N} blocks of graph-transformer-like architecture with a residual connected graph attention (GAT) layer and a feedforward neural network (FFN) layer.}
\label{fig:nqsqe}
\end{figure} 

\section{The NQS-gGA formalism}

In this section, we present the technical details of the NQS-gGA formalism developed in this work. As illustrated in Fig.\ref{fig:nqsqe} (a), the impurity model, generated by the QE outer loops, will be fed into the NQS solving process, which consists of the NQS optimization and properties sampling. For each step in the solving process, we have an error control metric to ensure the variance is minimized. In the following subsections, we will discuss the NQS method we used to construct the impurity solver in part A, the error control system that guarantees the stability of solving iterations in part B. In part C we will outline the gGA algorithmic structure.

\subsection{Neural Quantum States Impurity Solver}

The neural quantum state acts as a many-body wave function ansatz in the variational quantum Monte Carlo framework. Specifically, we consider an NQS wave function, expressed in second quantization as follows:
\begin{equation}
\begin{aligned}
    |\Psi_\theta\rangle&=\sum_x\langle x|\Psi_\theta\rangle|x\rangle=\sum_x\Psi_\theta(x)|x\rangle.
\label{eq:nqs1}
\end{aligned}
\end{equation}
Here $|x\rangle$ are Fock states, where $x$ encodes the corresponding spin-orbital occupation numbers, while $\Psi_\theta$ is a neural network, which takes an electronic configuration $x$ as input and outputs the corresponding wave-function amplitude $\Psi_\theta(x)$.

Using NQS, the expectation of any operator $\hat{O}$ is evaluated with Monte Carlo sampling as follows \cite{becca2017quantum}:
\begin{equation}
\begin{aligned}
    \langle \hat{O}\rangle=\mathbb{E}_{x\sim \Psi_\theta^2}\left[O_{loc}\right],\\
    O_{loc}=\frac{\langle x|\hat{O}|\Psi_\theta\rangle}{\langle x|\Psi_\theta\rangle},
\label{eq:nqs2}
\end{aligned}
\end{equation}
where $\mathbb{E}_{x\sim \Psi_\theta^2}$ represents the expectation over the probability distribution of the current parameterized wave function $\Psi_\theta$. We employ the Markov chain Monte Carlo \cite{brooks1998markov} to perform sampling and the neighbor-exchange method \cite{becca2017quantum} to generate new electron configurations. To approach the ground state, the NQS is optimized using stochastic reconfiguration \cite{sorella1998green} based on a recently proposed MinSR \cite{chen2024empowering} and the SPRING \cite{GOLDSHLAGER2024113351} methods to stabilise and accelerate the calculation.

Impurity models can be constructed with distinct geometries, such as linear chains or star-like structures \cite{lu2014efficient,lu2019natural}, as shown in Fig.~\ref{fig:model}, which arise from different ways of representing the non-interacting bath. This flexibility requires a network architecture capable of handling arbitrary hopping patterns between orbitals.
While transformer-based models effectively capture global interactions via attention mechanisms \cite{shang2023solving}, their computational cost scales as $O(N^3)$, where $N$ is the number of sites. Leveraging the sparse connectivity inherent to impurity models, we instead adopt the graph convolutional network ansatz. This reduces computation cost while maintaining the accuracy with an advanced network design.

More specifically, our graph-based model treats the orbital as graph nodes and single-particle interactions as edges. As illustrated in Fig.\ref{fig:nqsqe} (b), we initialise the node message as the orbital occupation number vector, which is a binary vector of value 1/0. Then, the state vectors are passed into the occupation positional encoding layer, where four different occupations in one site ($|11\rangle,|01\rangle,|10\rangle,|00\rangle$) are mapped to different vector embeddings, as is commonly adopted in natural language processing~\cite{vaswani2017attention}. Meanwhile, the positional information is added to the embedding vectors via the positional embeddings~\cite{vaswani2017attention} to uniquely distinguish each orbital site, which results in our final embeddings with all necessary state information. Furthermore, the embeddings are passed through the main body of the model, backboned by the GATv2~\cite{brody2021attentive} graph attention architecture. We mimic the transformer architecture by composing each iterative layer using one graph attention (GAT) layer and a feedforward neural network (FFN) layer with residual connections. After iterative accumulation, the features are then mapped via a linear layer to two real scalar quantities, corresponding to the wave function amplitude and the complex phases, as our wave function's output $\langle x|\Psi_\theta\rangle.$

A significant challenge in building a stable solver is to define a clear error metric for the wave function optimization. Conventional VMC provides an intuitive guideline — that lower variance typically correlates with reduced error, but the relationship is not quantitative. We will discuss this important issue in the following section.

\subsection{Error Control System}

A characteristic functional distinction between solving standalone Hamiltonians and acting as solvers to form self-consistent solutions is the requirement for numerical stability. In QE frameworks, the impurity model and the physical system are self-consistently coupled. The solution to the impurity model directly influences the state of the original system, which, in turn, affects the updates of the impurity problem itself. Uncontrolled errors can propagate through iterative loops, causing instability in the embedding procedure. Consequently, establishing a systematic error control methodology for the NQS solver within the variational Monte Carlo (VMC) framework is essential.

The errors arising in the VMC-NQS process can be categorized into two primary sources: (1) the wave function optimization error; (2) the Monte Carlo sampling error. The wave function optimization error refers to the discrepancy between the converged NQS wave function and the true ground-state wave function, which appears during the NQS optimization steps in Fig.\ref{fig:nqsqe} (a).

The quality of the optimized NQS wave function, $|\Psi_\theta\rangle$, is determined by how closely it approximates a true eigenstate of the impurity Hamiltonian. The most direct measure of this is the energy variance, $\text{Var}[E]$, which must vanish for an exact eigenstate and is therefore the fundamental diagnostic monitored during optimization:

\begin{equation}
\begin{aligned}
    \text{Var}[E]&=\mathbb{E}\left[H_{loc}^2\right] - \mathbb{E}^2\left[H_{loc}\right], \\
    H_{loc}&=\frac{\langle x|\hat{H}|\Psi_\theta\rangle}{\langle x|\Psi_\theta\rangle}.
\label{eq:ec1}
\end{aligned}
\end{equation}
However, $\text{Var}[E]$ depends on the system size and the overall energy scale of the Hamiltonian, which is an indirect measure of the true energy difference. To address this problem, here we adopt the so-called V-score~\cite{wu2024variational}:
\begin{equation}
\begin{aligned}
    V&=\frac{N\text{Var}[E]}{(E-E_T)^2},
\label{eq:ec2}
\end{aligned}
\end{equation}
which is a more robust, dimensionless, and size-independent metric of variational accuracy. Here $N$ is the number of sites, and the denominator provides a normalizing energy scale, where $E$ is the variational energy and $E_T$ is a fixed reference energy. 

We obtain this reference energy by averaging the energies $\langle\Psi_\epsilon|\hat{H}|\Psi_\epsilon\rangle$ of several randomly-generated Slater determinants, $|\Psi_\epsilon\rangle$, each of which can be evaluated analytically using the Hartree-Fock formula at a negligible computational cost. This method avoids the explicit Monte Carlo sampling used in \cite{wu2024variational} when treating $|\Psi_\epsilon\rangle$ as a true random many-body state.

The second major source of error is the statistical uncertainty from the Monte Carlo sampling performed during both the NQS optimization and properties sampling steps (see Fig.~\ref{fig:nqsqe}(a)). The numerical expectation of the observable from the Monte Carlo sampling is distributed as a Gaussian variable, whose standard deviation scales inversely as the square root of the sample size $M$ \cite{becca2017quantum}, i.e. $\sigma\sim O(1/\sqrt{M})$. This brings two critical uncertainties in the NQS solver: 
\begin{enumerate}
    \item Firstly, when using the V-score as the wave function optimization metric, it carries intrinsic statistical uncertainty due to finite sampling. This uncertainty is addressed by applying a confidence interval correction proportional to $O(1/\sqrt{M})$ to the computed V-score; we denote the arrived metric score as \textbf{E-tol}. 
    \item Secondly, when sampling embedded physical properties from the optimized wave function, it introduces direct uncertainty into the QE loop. This uncertainty must be independently controlled, adhering strictly to the convergence criteria of the outer embedding iterations (see big blue arrows in Fig.\ref{fig:model}). To bound this error, we apply a stringent criterion, typically $3\sigma$, corresponding to three times the standard deviation of the sampling expectation. This metric should be set lower than the convergence requirements of the embedding method. We denote the arrived properties sampling error metric as \textbf{P-tol}.
\end{enumerate}
 Notably, this convergence of P-tol often dominates the computational cost. For example, to reach a 0.001 residual for the embedding algorithm, it requires the error bound $3\sigma$ to be smaller than $5\times 10^{-4}$, corresponding to a standard deviation of approximately $\sigma=1.7\times10^{-4}$. This precision necessitates sampling in the order of $10^{8}$ configurations (as $M\propto 1/\sigma^2$), often exceeding the computational resources required for the wave function optimization itself.

By integrating the P-tol and E-tol metrics, we systematically control both the wave function optimization and sampling errors. This guarantees the convergence stability of the NQS impurity solver. We will demonstrate the practical effectiveness of this combined error-control strategy in Section 3 below. 

\subsection{The gGA Algorithmic Structure}

In this work we apply the ghost-Gutzwiller approximation (gGA) to the single-orbital Anderson lattice model:
\begin{align}
\hat{H}
&=
\sum_{\langle i,j\rangle}\sum_{\sigma}
\big(t_{ij}+\delta_{ij}\tau_p\big)\,
\hat{p}^\dagger_{i\sigma}\hat{p}_{j\sigma}
\nonumber
\\
&\quad+
V\sum_{i\sigma}
\Big(\hat{p}^\dagger_{i\sigma}\hat{d}_{i\sigma}
+\text{H.c.}\Big)
\nonumber
\\
&\quad+
\mu\sum_i\hat{N}_i
+\sum_i \hat{H}_{\mathrm{loc}}^i,
\label{eq:ex1}
\\
\hat{H}_{\mathrm{loc}}^i
&=
\frac{U}{2}\left(\hat{n}_{di}-1\right)^2.
\label{eq:ex1_loc}
\end{align}
Here $\{\hat{p}_{i\sigma}\}$ denote the itinerant (uncorrelated) degrees of freedom and $\{\hat{d}_{i\sigma}\}$ the interacting states, $\hat{n}_{di}=\sum_\sigma \hat{d}^\dagger_{i\sigma}\hat{d}_{i\sigma}$, and $\hat{N}_i=\hat{n}_{pi}+\hat{n}_{di}$ with $\hat{n}_{pi}=\sum_\sigma \hat{p}^\dagger_{i\sigma}\hat{p}_{i\sigma}$. The term $\hat{H}_{\mathrm{loc}}^i$ is the Hamiltonian of the $i$-th fragment.

The gGA is a variational framework based on the ansatz $|\Psi_G\rangle=\hat{\mathcal P}_G|\Psi_0\rangle$, where $|\Psi_0\rangle$ is a Slater determinant defined in an enlarged auxiliary Fock space and $\hat{\mathcal P}_G=\prod_i \hat{\mathcal P}_i$ is a product of local embedding maps into the physical Hilbert space~\cite{lanata2017emergent,frank2021quantum}. 
The variational energy is evaluated within the so-called Gutzwiller approximation~\cite{Gutzwiller3}, which becomes exact in the infinite-coordination-number limit.

As explained in Appendix, the gGA energy minimization problem can be cast in the form of a quantum-embedding (QE) framework~\cite{lanata2017emergent,frank2021quantum}.
In particular, for single-band models such as the ALM, the resulting algorithmic structure requires computing iteratively the ground state of an EH, consisting of the local fragment Hamiltonian $\hat{H}_{\mathrm{loc}}^i$ coupled to a finite bath of $B$ auxiliary orbitals:
\begin{align}
\hat{H}^{i}_{\mathrm{emb}}
&=
\hat{H}_{\mathrm{loc}}^i
\nonumber\\
&+
\sum_{a=1}^{B}\sum_{\sigma}
\Big([\mathcal{D}_i]_{a}\,\hat{d}_{i\sigma}^\dagger \hat{b}_{ia\sigma}
+\text{H.c.}\Big)
\nonumber\\
&+
\sum_{a,b=1}^{B}\sum_{\sigma}
[\Lambda_i^c]_{ab}\,\hat{b}_{ib\sigma}\hat{b}^\dagger_{ia\sigma}
\,,
\label{eq:gga4}
\end{align}
where $\{\hat{b}_{ia\sigma}\}$ are bath modes and $(\mathcal{D}_i,\Lambda_i^c)$ are updated self-consistently. The parameter $B$ controls the size of the variational space and, as shown in Refs.~\cite{lanata2017emergent,frank2021quantum,lee2023accuracy,lee2023accm}, setting $B=3$ is generally sufficient to calculate ground-state properties to DMFT-level accuracy.

A key advantageous property of gGA is that the self-consistency conditions require calculating only the expectation values of the following operators with respect to the ground state $\ket{\Phi_i}$ of $\hat{H}^{i}_{\mathrm{emb}}$:
\begin{align}
d_{ab\sigma}&=\langle \Phi_i| \hat{b}_{ib\sigma}\hat{b}^\dagger_{ia\sigma}|\Phi_i\rangle
\\
h_{b\sigma}&=\langle \Phi_i| \hat{d}^\dagger_{i\sigma}\hat{b}_{ib\sigma}|\Phi_i\rangle
\,.
\end{align}
This operation is the computational bottleneck of the gGA workflow, and motivates the development of scalable impurity solvers such as the NQS approach introduced here.

\subsubsection*{Symmetry Constraints}

In Sec.~\ref{Sec:benchmarks} we report gGA calculations performed enforcing spin-rotational invariance at the level of the variational ansatz. As shown in previous work~\cite{lee2024charge}, this requirement can be formulated in the quantum-embedding representation by demanding that the embedding problem (fragment plus bath) is treated in a manner consistent with global SU(2) spin rotations, i.e., by selecting the singlet solution of the EH. In practice, rather than implementing an explicit restriction to the $S^2=0$ sector, we enforce this condition by adding a positive penalty term to the EH,
\begin{equation}
\hat{H}^{i}_{\mathrm{emb}}
\ \longrightarrow\
\hat{H}^{i}_{\mathrm{emb}}+\alpha\,\hat{\mathbf S}^2,
\qquad
\alpha>0
\,,
\end{equation}
where $\hat{\mathbf S}$ is the total spin operator of the full embedding problem (including both fragment and bath). This shifts non-singlet states to higher energy and stabilizes the selection of the singlet ground state.
In addition, to reduce statistical noise and improve numerical stability, we apply a post-processing symmetrization of the spin structure of the one-body density matrices that enter the gGA self-consistency update.

As shown in Refs.~\cite{lanata2017emergent, frank2021quantum}, the symmetry condition that the gGA wavefunction is particle-number conserving (normal-state) is enforced by restricting the EH to the half-filled sector of the corresponding impurity$+$bath Hilbert space.
Within our NQS impurity solver, this is enforced by restricting the Monte Carlo sampling to configurations with the corresponding total particle number (number-conserving updates).

\begin{figure*}[htbp!]
\includegraphics[width=18 cm]{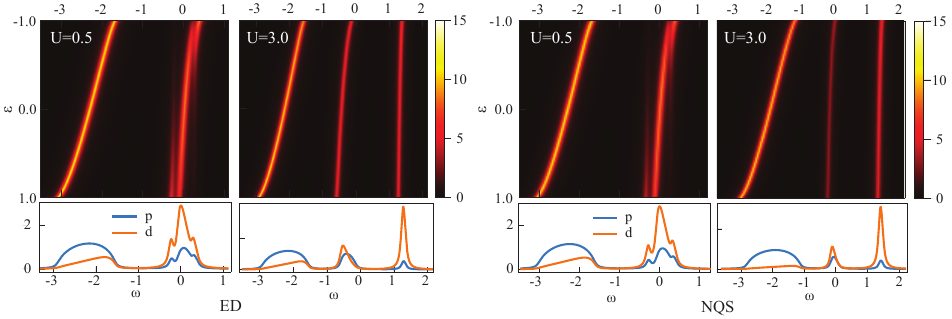}
\caption{The comparison of the solution of the Anderson lattice model using ED and NQS, combining with the gGA QE method. The band spectrum and density of states (DOS) are displayed. The convergence criteria are 1e-3 for both methods. The NQS impurity solver reproduces the phase difference correctly.}
\label{fig:nqs1o}
\end{figure*} 

\bigskip

\section{Benchmark Calculations of the ALM}
\label{Sec:benchmarks}

In this Section we present gGA benchmark calculations on the single-orbital ALM, see Eq.~\eqref{eq:ex1}, which exhibits a rich interplay between Mott physics and the hybridization between correlated and itinerant degrees of freedom, controlled by the on-site energy shift $\tau_p$ and the Hubbard interaction $U$. 

\paragraph{The Bethe lattice}

We focus on a Bethe lattice, within the limit of infinite coordination number, whose hopping-matrix eigenvalues have a semicircular free density of states:
\begin{equation}
    \rho(\epsilon)=\frac{2\sqrt{D^2-\epsilon^2}}{\pi D^2}
    \,.
\end{equation}

The spectral function of the ALM can be expressed as a function of the energy variable $\epsilon$, as follows:
\begin{equation}
\mathcal{A}(\epsilon,\omega)
=
\lim_{\eta\rightarrow 0^+}
-\frac{1}{\pi}\,\mathrm{Im}\,\mathrm{Tr}\!\left[\mathcal{G}(\epsilon,\omega+i\eta)\right]
\,,
\label{eq:ex2-a}
\end{equation}
where $\mathcal{G}$ is the Green's function, which can be calculated within the gGA formalism as discussed in the Appendix. In practice, we plot the broadened quantity obtained by evaluating the right-hand side at a small but finite $\eta$; in Fig.~\ref{fig:nqs1o} we use $\eta=0.06$ (in units of the half-bandwidth).
The corresponding density of states is obtained by averaging over $\epsilon$:
\begin{equation}
A(\omega)=\int d\epsilon\,\rho(\epsilon)\,\mathcal{A}(\epsilon,\omega)\,.
\label{eq:ex2}
\end{equation}
We refer the reader to Appendix for the explicit expression of $\mathcal{G}$ in terms of the converged gGA variational parameters.

\subsection{Benchmark calculations}

Throughout this section we set the half-bandwidth $D$ as the energy unit, $\tau_p=-1$, and consider two physical regimes: $U=0.5$ (metallic phase) and $U=3$ (Mott insulating phase).

We evaluate the performance of our NQS-based impurity solver using two accuracy settings: the default setting (NQS), with E-tol=$10^{-4}$ and P-tol=$10^{-4}$, and a higher-accuracy setting (NQS(hac)), with E-tol=$10^{-4}$ and P-tol=$5\times10^{-5}$.

As a first benchmark, we compare the spectral functions obtained with the higher-accuracy NQS setting in Fig.~\ref{fig:nqs1o}. 
At $U=0.5$, the spectral function shows metallic behaviour with finite density at the Fermi level, while at $U=3.0$ a gap opens, characteristic of a Mott insulator. By comparison with ED, the NQS solver captures the qualitative differences between these two regimes.

We next quantify the agreement through the occupation numbers reported in Table~\ref{tab:num_comparison}. In the metallic regime ($U=0.5$), both impurity ($d$) and bath ($p$) occupations are reproduced accurately by the NQS solver, with differences on the order of $10^{-3}$ relative to ED. In the insulating regime ($U=3$), both ED and NQS yield an almost fully occupied $p$ orbital, consistent with the insulating gap. However, the impurity occupation $n_d$ is more sensitive to numerical accuracy in this regime: the default NQS setting shows a larger deviation/uncertainty compared to ED, while the higher-accuracy setting NQS(hac) yields improved agreement (Table~\ref{tab:num_comparison}). At a high level, this increased sensitivity in the Mott regime can be traced to the self-consistent structure of quantum-embedding updates, where small solver inaccuracies can propagate more strongly than in the metallic regime; a brief discussion of this mechanism is provided in Appendix.

\begin{table*}[!hbt]
\centering
\setlength{\tabcolsep}{9pt}
\caption{The orbital occupation number computed at $U=0.5$ and $U=3.0$ by exact diagonalization (ED) and neural quantum state (NQS) solver. NQS(hac) stands for higher-accuracy, in which we use stricter error control E-tol and P-tol during the solve.}
\begin{tabular}{c|cc|cc}
\toprule
\multirow{2}{*}{Method} & \multicolumn{2}{c|}{$U = 0.5$} & \multicolumn{2}{c}{$U = 3.0$} \\
                        & $\mathbf{n}_d$ & $\mathbf{n}_p$ & $\mathbf{n}_d$ & $\mathbf{n}_p$ \\
\midrule
NQS & 0.6657$\pm$0.0006 & 0.8347$\pm$0.0004  & 0.5897$\pm$0.0480 & 0.9284$\pm$0.0051 \\
NQS(hac) & 0.6656 & 0.8342 & 0.5569 & 0.9355 \\
ED  & 0.6636 & 0.8348  & 0.5589 & 0.9348 \\
\bottomrule
\end{tabular}
\label{tab:num_comparison}
\end{table*}

Overall, our results showcase the effectiveness and suitability of the NQS solver within the gGA QE framework. Specifically, for gGA, the favorable computational scaling properties of NQS suggest potential for extending calculations to impurity models with larger orbital counts, potentially paving the way for applications to multi-orbital complex strongly correlated material systems beyond the reach of other methods.

\subsection{Error control analysis.}

\begin{figure*}[htbp!]
\includegraphics[width=16 cm]{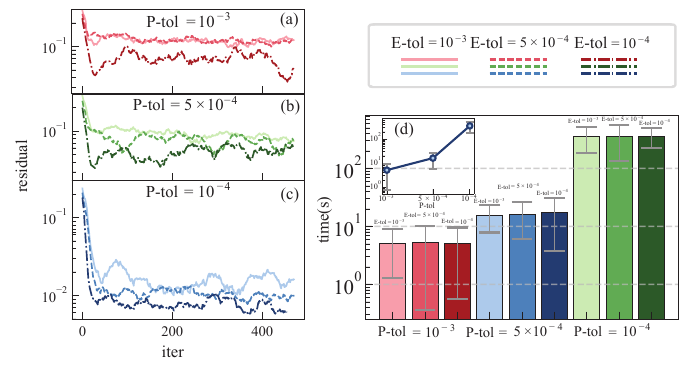}
\caption{Computational time cost and error analysis. (a)--(c) Residual during the self-consistent iterations for three sampling-accuracy settings:
(a) P-tol$=10^{-3}$, (b) P-tol$=5\times 10^{-4}$, and (c) P-tol$=10^{-4}$. Within each panel, the three curves correspond to different optimization tolerances: E-tol$=10^{-3}$ (solid), E-tol$=5\times 10^{-4}$ (dashed), and E-tol$=10^{-4}$ (dash-dotted). (d) Average wall time per property-sampling step for the same P-tol values; within each P-tol group, bars correspond to
E-tol$=10^{-3}$, $5\times 10^{-4}$, and $10^{-4}$ (as labeled above the bars). Error bars denote the standard deviation over the full set of iterations (400 solves).}
\label{fig:terr}
\end{figure*}

In this section, we present the error control metric E-tol and P-tol by comparing the convergence and time cost during independent gGA QE loops using the NQS impurity solver. In this analysis, we adopted $U=0.5$ metal phase Anderson lattice model discussed in the last section as a running example. We set the bounds for error control metrics P-tol and E-tol to be among ($10^{-3}$, $5\times10^{-4}$, $10^{-4}$). All other settings, including the network structure, sizes, and the VMC optimization parameters, are kept consistent in all tests. The analysis results are displayed in Fig.\ref{fig:terr}. From this, we verified the joint importance of both E-tol and P-tol. We also observe considerable growth of the time cost when reducing the P-tol.

First, we display the residual computed from the maximum absolute difference of the gGA parameters $\mathcal{R}_i$ and $ \Lambda_i$ between the previous and current iterations. The results are arranged in subfigures (a)-(c) of Fig.\ref{fig:terr}. In all cases, the residual decays rapidly in the first several iterations, followed by long and slow convergence to smaller values. The decrease of $\text{E-tol}$ can assist in reducing the residual: for instance, for $\text{E-tol}$=$10^{-4}$ (dark red line), lower residual results, compared to other cases when fixing the $\text{P-tol}$. Since this generally appears, we therefore conclude that the E-tol, which represents the wave function optimization error, is important to the stability of the NQS VMC impurity solver. Moreover, comparing the three subfigures, we notice that the residual cannot be further reduced when $\text{P-tol}$ is larger than $10^{-4}$, regardless of the choice of the bound for E-tol. For instance, in Fig.~\ref{fig:terr}(a) with $\text{P-tol}$ = $10^{-3}$, it appears that all samples cannot converge to a considerably smaller residual. This is caused by the instabilities of sampled properties, which lead to unstable quantum-embedding loops. While for $\text{P-tol}$=$10^{-4}$ (blue lines in Fig.~\ref{fig:terr} (c)), the convergence of all three samples with different $\text{E-tol}$ is improved. This shows the importance of controlling the property sampling accuracy P-tol, in our method. 

Nevertheless, a higher accuracy of the properties (smaller P-tol) comes with considerable computational cost. In Fig.~\ref{fig:terr} (d), we displayed the time cost analysis of the properties sampling. We keep the colour convention consistent with (a)-(c), where lighter ones indicate higher $\text{E-tol}$ and darker ones indicate smaller $\text{E-tol}$. The time costs are computed by averaging the cost of sampling steps in each solving iteration. An average of more than 400 iterations is performed, which gives standard deviation as illustrated by the error bars in Fig.~\ref{fig:terr} (d). We can clearly see the growth of the sampling time with the decrease of the bound of P-tol. We observe the time cost increases by over 100 times from $\text{P-tol}$=$10^{-3}$ to $10^{-4}$, while the increase is around 10 times to $\text{P-tol}$=$5\times10^{-4}$. This aligns with the scaling of the VMC uncertainties $\epsilon$, where $\epsilon\propto O(1/\sqrt{M})$. In this case, a decreased error tolerance by an order of magnitude would increase $M$ by about 100 times. Comparably, the time cost for the wave function optimization appears negligible, which only takes (1.63, 2.31, 5.70) seconds per optimization step on average for $\text{E-tol}$ among ($10^{-3}$, $5\times10^{-4}$, $10^{-4}$). From this, we also observe that the growth of computational time with the decrease of E-tol is rather gentle, compared with that of the parameter P-tol.

To conclude, the error analysis demonstrates that E-tol and P-tol are both important metrics for a numerically accurate solution. The dominant computational time cost is the properties sampling rather than the wave function optimization. This finding calls for a better VMC sampling method to be developed for quickly extracting the accurate physical observables from the NQS.

\section{Discussion and summary}

Employing NQS as an impurity solver offers several advantages. Due to the universal approximation capability of neural networks, accurately approximating the ground state wave function is feasible with properly designed network architectures, significantly reducing approximation errors common in other solvers. Furthermore, the favourable computational scaling properties of variational Monte Carlo (VMC) enable straightforward extensions to multi-orbital models relevant for both theoretical investigation and realistic material simulation. Typically, the neural network evaluation scales as $O(N^2)$, while the Monte Carlo sampling scales as $O(N)$ \cite{chen2025neural}. With proper parallelisation on modern computers, such scalability suggests significant potential for studying complex systems such as strongly interacting f-orbital materials and the extended active-space models. This provides valuable possibilities for advancing research on strongly correlated electron systems.

Our study indicates that there are remaining challenges that need to be solved for further developing the most robust and general-purpose NQS impurity solver. 
\begin{itemize}
    \item \textbf{Accelerated computation of embedding properties.}
    Iterative updates in QE algorithms require precise computation of various physical observables. For example, the DMFT requires the Green's Function as the matching target, while the DMET and the gGA rely on accurate evaluation of single-body and/or two-body reduced density matrices. In practice, the sampling accuracy of these properties strongly affects the convergence of the entire embedding algorithm. Insufficient sampling accuracy can result in violations of fundamental physical constraints, therefore potentially destabilising the iterative embedding cycle, as found in our analysis. Given the slow $O(1/\sqrt{M})$ convergence of the sampling accuracy, a naive sampling strategy requires very large computational resources - for instance, roughly $10^8$ samples to reach an accuracy of $10^{-3}$. Therefore, developing an enhanced or importance-sampling method tailored specifically to NQS-VMC represents a useful research direction for enhancing computational efficiency and broadening the applicability of the NQS method. In addition, it would also be useful to benchmark the stability and numerical accuracy requirements of different QE frameworks. For instance, workflows such as NQS-DMET and NQS-DMFT could have different requirements for sampling accuracy.
    \item \textbf{Incorporating spin symmetries in NQS wave function.}
    In strongly correlated electronic systems, it is often necessary to enforce specific spin symmetries (e.g., spin degeneracy, collinear, non-collinear spin configurations, etc). However, embedding these symmetries explicitly into NQS wave functions is nontrivial. Incorrect symmetries could lead to convergence issues and a physically incorrect solution. One practice, employed in this study, involves modifying the Hamiltonian with spin symmetry-related penalties and post-hoc symmetrisation. While robust in this study, it may also be beneficial to incorporate the symmetry inherent in the wave function ansatz. This may require symmetry-aware network structure design.
    \item \textbf{Exploring multiple t-U region and larger system.}
    While the focus of the current work is to identify sampling issues and establish robust convergence criteria for NQS within a quantum-embedding framework, it is useful to apply it for larger systems and scan a wider range of interacting and hopping parameters.
\end{itemize}

In summary, we have designed and benchmarked an NQS-based impurity solver with an error control system. When integrated with the ghost Gutzwiller Approximation on the Anderson Lattice Model, our solver accurately captures the key electronic properties of both its metallic and Mott insulating phases, yielding results in excellent agreement with exact diagonalization. Importantly, our analysis of the computational budget identifies the primary bottleneck not as the wave function optimization, but as the high-accuracy property sampling required by the embedding loop. This finding, therefore, establishes the development of efficient, high-precision sampling methods as the primary challenge to establishing NQS as a practical impurity solver within QE methods.

\section*{Code and data availability}
The full data and the code to reproduce the results of this work are publicly available on the GitHub website. The website link is: (https://github.com/floatingCatty/Hubbard). For the ED solver, we employ the QuSpin packages (https://quspin.github.io/QuSpin/). The NQS is written as an extension to the advanced VMC package quantax (https://github.com/ChenAo-Phys/quantax).

\begin{acknowledgments}
We gratefully acknowledge the financial support of the Natural Science and Engineering Research Council of Canada (Z.Z. and H.G.). H.G. wishes to thank Prof. Xi Dai for useful discussions on the 
Gutzwiller formalism. T.-H.L. gratefully acknowledges funding from the National Science and Technology Council (NSTC) of Taiwan under Grant No. NSTC 112-2112-M-194-007-MY3 and the National Center for Theoretical Sciences (NCTS) in Taiwan. N.L. acknowledges support from the National Science Foundation under Award No. DMR-2532771. We thank the Digital Research Alliance Canada for providing substantial computational facilities, which made this work possible. 
\end{acknowledgments}

\appendix

\section{Details of the Ghost Gutzwiller algorithm}
\label{app:gGA_algorithm}

For completeness, here we outline the standard ghost-Gutzwiller approximation (gGA) equations in the quantum-embedding form, as derived in Refs.~\cite{lanata2017emergent,frank2021quantum}.

\subsubsection{Quantum-embedding Lagrange formulation of gGA}

We consider a generic multi-orbital Hubbard Hamiltonian written in fragment form
\begin{align}
\hat{H}
&=
\sum_{\substack{i,j=1\\ i\neq j}}^{\mathcal{N}}
\sum_{\alpha=1}^{\nu_i}\sum_{\beta=1}^{\nu_j}
[t_{ij}]_{\alpha\beta}\,\hat{c}^\dagger_{i\alpha}\hat{c}_{j\beta}
+\sum_{i=1}^{\mathcal{N}}
\hat{H}^{i}_{\mathrm{loc}}\!\left[\hat{c}^\dagger_{i\alpha},\hat{c}_{i\alpha}\right].
\label{eq:H_app}
\end{align}
Here the inter-fragment hopping terms are collected in the $t_{ij}$ blocks ($i\neq j$), while $\hat{H}^{i}_{\mathrm{loc}}$ contains all local contributions on fragment $i$, including both the on-site one-body terms and the local interaction terms.

The gGA is a variational framework based on the ansatz
\begin{align}
|\Psi_G\rangle
&=
\hat{\mathcal{P}}_G\,|\Psi_0\rangle,
\label{eq:gGA_ansatz_app}
\end{align}
where $|\Psi_0\rangle$ is a single-particle wavefunction and $\hat{\mathcal{P}}_G$ is an operator, both to be variationally determined. A key feature of gGA is that, differently from the standard GA, $|\Psi_0\rangle$ is constructed in an auxiliary Fock space of tunable dimension. Specifically, for each fragment $i$ one introduces $B\nu_i$ auxiliary (ghost) fermionic modes $\hat{f}_{ia}$ ($a=1,\dots,B\nu_i$), where $B$ is a positive integer. The operator $\hat{\mathcal{P}}_G$ has a local-product structure $\hat{\mathcal{P}}_G=\prod_{i=1}^{\mathcal{N}}\hat{\mathcal{P}}_i$, where each $\hat{\mathcal{P}}_i$ is a local embedding map from the auxiliary local Fock space (generated by the $\hat{f}_{ia}$) to the physical local Fock space (generated by the $\hat{c}_{i\alpha}$). The variational energy is
\begin{align}
\mathcal{E}
&=
\langle\Psi_G|\hat{H}|\Psi_G\rangle
=
\langle\Psi_0|\hat{\mathcal{P}}_G^\dagger\,\hat{H}\,\hat{\mathcal{P}}_G|\Psi_0\rangle,
\end{align}
to be minimized with respect to $|\Psi_0\rangle$ and $\hat{\mathcal{P}}_G$.

In practice, the evaluation of $\mathcal{E}$ is performed by imposing the standard Gutzwiller constraints and adopting the Gutzwiller approximation, i.e., neglecting the Wick-contraction contributions that vanish in the infinite-coordination-number limit~\cite{lanata2017emergent,frank2021quantum}. Within this framework, the variational energy-minimization problem can be formulated as the stationarity of the following Lagrange function~\cite{lanata2017emergent,frank2021quantum}:
\begin{align}
\mathcal{L}
&=
\langle \Psi_0 | \hat{H}_{\mathrm{qp}}[\{\mathcal{R}_i\},\{\Lambda_i\}] | \Psi_0 \rangle
+E\left(1-\langle \Psi_0 | \Psi_0 \rangle\right)
\nonumber\\
&\quad+
\sum_{i=1}^{\mathcal{N}}
\Big[
\langle \Phi_i | \hat{H}^{i}_{\mathrm{emb}}[\mathcal{D}_i,\Lambda_i^c] | \Phi_i \rangle
+E^c_i\left(1-\langle \Phi_i | \Phi_i \rangle\right)
\Big]
\nonumber\\
&\quad-
\sum_{i=1}^{\mathcal{N}}
\sum_{a,b=1}^{B\nu_i}
\Big(
[\Lambda_i]_{ab}+[\Lambda_i^c]_{ab}
\Big)
[\Delta_i]_{ab}
\nonumber\\
&\quad-
\sum_{i=1}^{\mathcal{N}}
\sum_{a,b=1}^{B\nu_i}\sum_{\alpha=1}^{\nu_i}
\Big(
[\mathcal{D}_i]_{a\alpha}\,[\mathcal{R}_i]_{b\alpha}\,
\big[\Delta_i(\mathbf{1}-\Delta_i)\big]^{\tfrac{1}{2}}_{ba}
+\mathrm{c.c.}
\Big).
\label{eq:L_old_app}
\end{align}
The following matrices of Lagrange multipliers satisfy:
$\mathcal{R}_i,\mathcal{D}_i
\in\mathbb{C}^{B\nu_i\times\nu_i}$,
$\Lambda_i=\Lambda_i^\dagger$,
$\Lambda_i^c=(\Lambda_i^c)^\dagger$,
$\Delta_i=\Delta_i^\dagger
\in\mathbb{C}^{B\nu_i\times B\nu_i}$, while
the scalar Lagrange multipliers $E,E_i^c$ enforce normalization of $|\Psi_0\rangle$ and $|\Phi_i\rangle$, respectively.

The auxiliary Hamiltonians entering Eq.~\eqref{eq:L_old_app} are
\begin{align}
\hat{H}_{\mathrm{qp}}[\{\mathcal{R}_i\},\{\Lambda_i\}]
&=
\sum_{\substack{i,j=1\\ i\neq j}}^{\mathcal{N}}
\sum_{a=1}^{B\nu_i}\sum_{b=1}^{B\nu_j}
\big[\mathcal{R}_i\,t_{ij}\,\mathcal{R}_j^\dagger\big]_{ab}\,
\hat{f}^\dagger_{ia}\hat{f}_{jb}
\nonumber\\
&\quad+
\sum_{i=1}^{\mathcal{N}}
\sum_{a,b=1}^{B\nu_i}
[\Lambda_i]_{ab}\,\hat{f}^\dagger_{ia}\hat{f}_{ib},
\label{eq:Hqp_app}
\\
\hat{H}^{i}_{\mathrm{emb}}[\mathcal{D}_i,\Lambda_i^c]
&=
\hat{H}^{i}_{\mathrm{loc}}\!\left[\hat{c}^\dagger_{i\alpha},\hat{c}_{i\alpha}\right]
\nonumber\\
&\quad+
\sum_{\alpha=1}^{\nu_i}\sum_{a=1}^{B\nu_i}
\Big(
[\mathcal{D}_i]_{a\alpha}\,\hat{c}^\dagger_{i\alpha}\hat{b}_{ia}
+\mathrm{H.c.}
\Big)
\nonumber\\
&\quad+
\sum_{a,b=1}^{B\nu_i}
[\Lambda_i^c]_{ab}\,\hat{b}_{ib}\hat{b}^\dagger_{ia}.
\label{eq:Hemb_app}
\end{align}

For later use, we define the following block matrices (with vanishing diagonal blocks for $t$):
\begin{align}
t
&=
\begin{pmatrix}
\mathbf{0} & t_{12} & \dots & t_{1\mathcal{N}}\\
t_{21} & \mathbf{0} & \dots & \vdots\\
\vdots & \vdots & \ddots & \vdots\\
t_{\mathcal{N}1} & \dots & \dots & \mathbf{0}
\end{pmatrix},
\label{eq:block_t_app}
\\
\mathcal{R}
&=
\begin{pmatrix}
\mathcal{R}_1 & \mathbf{0} & \dots & \mathbf{0}\\
\mathbf{0} & \mathcal{R}_2 & \dots & \mathbf{0}\\
\vdots & \vdots & \ddots & \vdots\\
\mathbf{0} & \mathbf{0} & \dots & \mathcal{R}_{\mathcal{N}}
\end{pmatrix},
\label{eq:block_R_app}
\\
\Lambda
&=
\begin{pmatrix}
\Lambda_1 & \mathbf{0} & \dots & \mathbf{0}\\
\mathbf{0} & \Lambda_2 & \dots & \vdots\\
\vdots & \vdots & \ddots & \vdots\\
\mathbf{0} & \dots & \dots & \Lambda_{\mathcal{N}}
\end{pmatrix},
\label{eq:block_Lambda_app}
\end{align}
so that the one-body matrix associated with $\hat{H}_{\mathrm{qp}}$ is
\begin{align}
h^*[\mathcal{R},\Lambda]
&=
\mathcal{R}\,t\,\mathcal{R}^\dagger+\Lambda
\,.
\label{eq:hstar_app}
\end{align}

\subsubsection{Stationarity equations}

To extremize the Lagrange function in Eq.~\eqref{eq:L_old_app}
we represent $\Delta_i$, $\Lambda_i$, and $\Lambda_i^c$ in an orthonormal basis of Hermitian matrices
$\{[h_i]_s\}_{s=1}^{(B\nu_i)^2}$ with respect to the canonical scalar product
\begin{align}
(A,B)=\mathrm{Tr}\!\left[A^\dagger B\right],
\label{eq:canonical_scalar_product_app}
\end{align}
as follows:
\begin{align}
\Delta_i
&=
\sum_{s=1}^{(B\nu_i)^2}\left[d_i\right]_s\,[h_i^{T}]_s,
\nonumber\\
\Lambda_i
&=
\sum_{s=1}^{(B\nu_i)^2}\left[l_i\right]_s\,[h_i]_s,
\nonumber\\
\Lambda_i^c
&=
\sum_{s=1}^{(B\nu_i)^2}\left[l_i^c\right]_s\,[h_i]_s,
\label{eq:basis_expansions_app}
\end{align}
with real-valued coefficients $\left[d_i\right]_s,\left[l_i\right]_s,\left[l_i^c\right]_s$.

With this notation, the gGA Lagrange equations can be cast as follows:
\begin{widetext}

\begin{align}
\hat{H}_{\mathrm{qp}}[\{\mathcal{R}_i\},\{\Lambda_i\}]\,|\Psi_0\rangle
&=
E_0\,|\Psi_0\rangle,
\label{eq:sp1_app}
\\
\hat{H}^{i}_{\mathrm{emb}}[\mathcal{D}_i,\Lambda_i^c]\,|\Phi_i\rangle
&=
E^{c}_i\,|\Phi_i\rangle,
\label{eq:sp2_app}
\\
[\Delta_i]_{ab}
&=
\langle\Psi_0|\hat{f}^\dagger_{ia}\hat{f}_{ib}|\Psi_0\rangle,
\label{eq:sp3_app}
\\
\sum_{c=1}^{B\nu_i}
[\mathcal{D}_i]_{c\alpha}\,
\big[\Delta_i(\mathbf{1}- \Delta_i)\big]^{\tfrac{1}{2}}_{ac}
&=
\sum_{\substack{j=1\\ j\neq i}}^{\mathcal{N}}
\sum_{\beta=1}^{\nu_j}
[t_{ij}]_{\alpha\beta}
\sum_{b=1}^{B\nu_j}
[\mathcal{R}_j^\dagger]_{\beta b}\,
\langle\Psi_0|\hat{f}^\dagger_{ia}\hat{f}_{jb}|\Psi_0\rangle,
\label{eq:sp4_app}
\\
\left[l^c_i\right]_{s}
&=
-\left[l_i\right]_{s}
-\sum_{c,b=1}^{B\nu_i}\sum_{\alpha=1}^{\nu_i}
\frac{\partial}{\partial \left[d_i\right]_s}
\Bigg(
\big[\Delta_i(\mathbf{1}-\Delta_i)\big]^{\tfrac{1}{2}}_{cb}\,
[\mathcal{D}_i]_{b\alpha}\,
[\mathcal{R}_i]_{c\alpha}
+\mathrm{c.c.}
\Bigg),
\label{eq:sp5_app}
\\
[\Delta_i]_{ab}
&=
\langle\Phi_i|\hat{b}_{ib}\hat{b}^\dagger_{ia}|\Phi_i\rangle,
\label{eq:sp6_app}
\\
\sum_{a=1}^{B\nu_i}
[\mathcal{R}_i]_{a\alpha}\,
\big[\Delta_i(\mathbf{1}-\Delta_i)\big]^{\tfrac{1}{2}}_{ab}
&=
\langle\Phi_i|\hat{c}^\dagger_{i\alpha}\hat{b}_{ib}|\Phi_i\rangle.
\label{eq:sp7_app}
\end{align}

\end{widetext}

\subsubsection{Self-consistent solution strategy}

Equations \eqref{eq:sp1_app}--\eqref{eq:sp7_app} are solved by fixed-point iteration. A convenient organization is the following cycle:
\begin{enumerate}
\item Starting from a guess for $\{\mathcal{R}_i,\Lambda_i\}$, solve the quadratic quasiparticle problem \eqref{eq:sp1_app} for $|\Psi_0\rangle$ and evaluate the equal-time one-body correlators entering \eqref{eq:sp3_app} and \eqref{eq:sp4_app}.
\item With $\{\mathcal{R}_i,\Lambda_i\}$ and $\{\Delta_i\}$ fixed, determine $\{\mathcal{D}_i,\Lambda_i^c\}$ from the stationarity conditions \eqref{eq:sp4_app} and \eqref{eq:sp5_app}.
\item With $\{\mathcal{D}_i,\Lambda_i^c\}$ fixed, solve the interacting EHs \eqref{eq:sp2_app} to obtain $\{|\Phi_i\rangle\}$.
\item Update $\{\Delta_i\}$ and $\{\mathcal{R}_i\}$ using the matching conditions \eqref{eq:sp6_app} and \eqref{eq:sp7_app}, and iterate until convergence.
\end{enumerate}
The computational bottleneck is the repeated solution of the interacting EHs
$\hat{H}^{i}_{\mathrm{emb}}[\mathcal{D}_i,\Lambda_i^c]$.

\subsubsection{Physical observables and Green's functions}

At convergence, local observables are obtained from the embedding ground states:
\begin{align}
\langle\Psi_G|\hat{O}^{i}_{\mathrm{loc}}|\Psi_G\rangle
&=
\langle\Phi_i|\hat{O}^{i}_{\mathrm{loc}}|\Phi_i\rangle.
\label{eq:obs_local_app}
\end{align}
For $i\neq j$, the gGA approximation yields
\begin{align}
\langle\Psi_G|\hat{c}^\dagger_{i\alpha}\hat{c}_{j\beta}|\Psi_G\rangle
&=
\sum_{a=1}^{B\nu_i}\sum_{b=1}^{B\nu_j}
[\mathcal{R}_i^\dagger]_{\alpha a}\,
\langle\Psi_0|\hat{f}^\dagger_{ia}\hat{f}_{jb}|\Psi_0\rangle\,
[\mathcal{R}_j]_{b\beta}.
\label{eq:obs_nonlocal_app}
\end{align}

The physical single-particle Green's function is expressed in terms of the converged variational parameters as follows:
\begin{align}
G(z)
&=
\mathcal{R}^\dagger\Big[z\mathbf{1}-h^*[\mathcal{R},\Lambda]\Big]^{-1}\mathcal{R},
\label{eq:G_app}
\end{align}
with $h^*[\mathcal{R},\Lambda]$ defined in Eq.~\eqref{eq:hstar_app}. The corresponding spectral function is
\begin{align}
\mathcal{A}(\omega)
&=\lim_{\eta\rightarrow 0^+}
-\frac{1}{\pi}\,\mathrm{Im}\,\mathrm{Tr}\,G(\omega+i\eta)
\,.
\label{eq:spectral_A_app}
\end{align}

\section{Evaluation of the reference mean-field energy $E_T$.}
As discussed in the main text, we choose the $E_T$ in the V-score computation:
\begin{equation}
V=\frac{N\text{Var}[E]}{(E-E_T)^2},
\end{equation}
as the energy expectation of a random mean-field wave function. Therefore, we can use the closed-form Hartree-Fock formula to estimate the energy accurately, reducing the need for sampling.

To construct the random mean-field solution with the physical constraint, we used a parameterisation technique. The closed-form Hartree-Fock energy expression is:
\begin{equation}
E_{hf}=\frac{1}{2}\text{Tr}[D\cdot h]+\frac{1}{2}\text{Tr}[D\cdot F],
\end{equation}
where the $h$ are the one-body terms of the original system Hamiltonian, and the $F$ is the corresponding Fock matrix.

Since this expression is related directly to the density matrix, we need to design a sampling method to ensure the density matrix respects the physical constraints, such as particle conservation and idempotency. To achieve this, we first generate a random hermitian matrix $A$, then, by performing the QR decomposition, we attain a set of orthogonal random basis vectors $Q$. Afterwards, by selecting $n$ columns of matrix $Q^{(n)}=Q[0:n]$ (n is the total electron occupation number), and multiplying by its conjugate transpose, we have the sampled form $\tilde{D}=Q^{(n)}{Q^{(n)}}^\dagger$. The matrix $\tilde{D}$ follows all the required properties of the physical mean-field density matrix. Later on, we use $\tilde{D}$ to construct the Fock matrix $F$, therefore compute the energy using the above formula. After averaging over sampled $\tilde{D}$ and $F$, the converged energy will be used as the reference $E_T$.

\section{Computational Details and Hyperparameters}
In this section, we detail the numerical setup in different sections of the NQS-gGA workflow. Including the NQS solver settings, the updating algorithm of gGA self-consistent loops, and the overall convergence criteria.
\subsubsection{NQS solver}
We adopt a two-stage workflow for the NQS solver to minimize the computational cost while maintaining consistency and robustness. First, the solver is initialized with the user-defined hyperparameters. These are: 
\begin{itemize}
    \item d\_emb=4 the embedding vector dimension on each orbital occupation, 
    \item Nsamples=5000 the number of MC samples per sweep,
    \item nblocks=3 the neural network depth, 
    \item hidden\_channels=8 the size of the network's hidden channels,
    \item out\_channels=8 the size of the network output channels,
    \item ffn\_hidden=[8] the size of the feed forward networks hidden channels,
    \item heads=4 the number of the attention heads. 
\end{itemize}

These parameters define the network structure. We also set the P-tol and E-tol here. We use E-tol=$10^{-4}$ and P-tol=$10^{-4}$ in the ALM benchmark solve.

After initialization, the solvers are ready to work. For each solving iteration, the model is separated into two phases. First, a mean-field wave function ansatz is used. The optimization reaches convergence for the mean-field solution first. Then, a neural network ansatz is added on the meanfield solution following the rules used in \cite{chen2025neural}, which is optimized with extra iterations to reach a predefined energy tolerance E-tol. The learning rate of all steps is fixed at 0.01, with the MinSR+SPRING\cite{chen2024empowering, GOLDSHLAGER2024113351} stochastic reconfiguration optimizer with pseudo-inverse threshold of $10^{-6}$ and SPRING damping factor of 0.5. 

We note that the EHs in the QE loops are varied continuously during iterative updates; therefore, the solution of the next will have a similar structure to the previous one. This motivates us to save the converged wave function after each solve and use it as an initialisation of the next. This reduced the optimization cost as the convergence of the outer QE loop is approached. However, as we reported in the experiment that the total cost is dominated by the sampling phase, the overall reduction in cost using this warm start method in wave function optimization is still considerably minor.

\subsubsection{Parameter update}
We use a mixing method in the gGA updates. In all experiments, we adopt a linear mixing with a mixing $\beta$ of 0.3. This is used for increasing the stability of the solution. In linear mixing, the new updated parameter will be averaged with the previous parameter by a factor $\beta$, where the updating formula is:
$$P=(1-\beta)P_{old}+\beta*P_{new}$$
Where P is the updated parameter, $P_{old}$ and $P_{new}$ are the parameters for the parameter used in the current iteration and the computed new parameter updates.

Besides, our package implements a periodic Pulay mixing method\cite{banerjee2016periodic}, which is a second-order mixing method and has good acceleration on convergence in many cases.

\subsubsection{Convergence criteria}
The convergence criteria are separated into three distinct levels. First, the NQS solver optimizes the ansatz wave function until the energy variance satisfies the E-tol threshold. Second, the properties sampling step proceeds, until the error of the sampled properties reaches P-tol. Specifically, the statistical uncertainty of each matrix element in the computed reduced density matrix must be smaller than the defined P-tol value. Finally, after solving the EH, the updated parameters from the QE algorithm are compared against those from the previous iteration. A threshold is applied to the maximum absolute error of the parameters to determine the convergence of the global QE workflow.

\subsubsection{Computational Platform}
We use devices from the Narval cluster node of the Digital Research Alliance of Canada for all computation. We use one NVidia A100SXM4 (40 GB memory) GPU for each solve, and 12 CPUs from AMD EPYC 7413 with 125 GB memory in total.

\bibliography{main.bib}
\end{document}